\documentclass[a4paper,fleqn]{scrartcl}
\usepackage[utf8]{inputenc}
\usepackage[T1]{fontenc}
\usepackage{lmodern}
\usepackage{graphicx}% Include figure files
\usepackage{dcolumn}% Align table columns on decimal point
\usepackage{bm}% bold math
\usepackage{amsmath}
\usepackage{caption}[2008/07/24]
\newcommand{\beq}{\begin{equation}}
\newcommand{\eeq}{\end{equation}}
\newcommand{\beqa}{\begin{eqnarray}}
\newcommand{\eeqa}{\end{eqnarray}}

\begin{document}

\begin{center}
%{{\Large\bf {Chiral symmetry, nuclear forces and all that}}}
{{\huge\bf{\sf {Chiral symmetry, nuclear forces and all that}}}}

\vspace{.4cm}

{\large Ulf-G. Mei\ss ner}

\bigskip

{\sl  HISKP and Bethe Center for Theoretical Physics\\
     Universit\"at Bonn, D-53115 Bonn, Germany\\and\\
     Institut f\"ur Kernphysik, Institute for Advanced Simulation and JCHP\\ 
     Forschungszentrum J\"ulich, D-52425 J\"ulich, Germany\\[0.9em]
        {\rm Dedicated to Gerry Brown on the occasion of his $85^{th}$ birthday}}

\bigskip

\begin{abstract}
These are personal recollections of  how Gerry Brown's work and thinking
influenced the development of the 
chiral effective Lagrangian based theory of nuclear forces.
\end{abstract}
\end{center}

\bigskip

\section{The early days at Stony Brook}

I first met Gerry on the occasion of the Erice school in 1981, where he
and Mannque Rho were giving quite interesting lectures on the chiral bag model
and related issues. At that point, I was close to finishing my diploma
thesis under the supervision of Manfred Gari at Bochum, working on two-meson
exchange currents in the framework of unitary transformations. More
specifically, the goal was to find out the relative contribution of the
isoscalar $\rho \pi^2 \gamma$ current compared to the well-established
leading order isoscalar  $\rho \pi \gamma$ current. This calculation turned
out to be - to use Howard Georgi's words - an excellent exercise in
self-torture, as one had to evaluate about 1000 time-ordered diagrams by hand. Also,
the result that these corrections were already sizeable at momentum transfers
below $q^2 \simeq 1\,$GeV$^{2}$ did not quite fit into the standard lore which was
so eloquently summarized by Gerry and Mannque in their chiral filter
hypothesis \cite{Rho:1981zi}. This hypothesis that grew out of explicit
meson-exchange current calculations including heavier hadrons like the
$\Delta(1232)$ or the vector mesons $\rho, \omega$ was clearly emphasizing
the role of the pion. Whenever present, the one-pion exchange current was
supposed to dominate the pertinent response of light nuclei
to electromagnetic probes, even up to momentum transfers as large as
1~GeV$^2$, way beyond the soft-pion limit. 
In fact, it was clear to Gerry and Mannque -- and others --
 that the understanding
of the nucleon structure from QCD and the nuclear interactions were two
intimately connected issues, firmly rooted in the chiral symmetry of
QCD. It is worth quoting from Ref.~\cite{Rho:1981zi}: ``... that there
must exist an intimate connection between what makes up a hadron and
what induces a rich variety of interactions between hadrons. ..., it will 
be rather unlikely one will gain a full understanding of one without the
other.'' Only then too much emphasis was put onto the quark structure of
the nucleon - as we understand now,  low-energy QCD is well approximated
by an hadronic effective Lagrangian. When doing my calculations at Bochum, 
I did not know that, but learned about it quickly when I arrived at 
Stony Brook in April of 1982. The Erice lectures on the little 
bag model had aroused my interest and, honestly, I was fed up
calculating meson-exchange currents. Therefore, I contacted Gerry  
about becoming a graduate student at Stony
Brook. It did not take him long to accept me - as it turned out, I spent
a short but very productive time in Gerry's group until my graduation in
December 1984. But let me briefly come back to the meson-exchange current
(MEC) story. In fact, only after some strong support from Dan-Olof Riska, one
of the MEC pioneers \cite{RB}, my diploma work was finally published
\cite{Meissner:1983wn} but largely ignored by the community. The crux of the 
matter was that in those days one did not have a power counting that allowed
one to systematically address the corrections to the leading order one-pion
contribution either in the nuclear forces or in the corresponding exchange 
currents, though a tremendous amount of phenomenology had been built up
over the years. In particular, the Stony Brook and Paris approach to the
two-nucleon force that used dispersion relations to connect the $\pi N$ 
scattering amplitudes to the $2\pi$-exchange in the NN force supplemented
by some heavy mass Yukawa-type functions with parameters determined from
a fit to the data where already very close to what would become later the
``modern theory of nuclear forces''. In any case, I had to learn the
basis of the fundamentals of the Stony Brook potential by working through
the book of Gerry and Andy Jackson \cite{BJbook}. This turned out to be 
a very painful exercise as the notation would change from chapter to chapter,
but made me familiar with dispersion theory and the many intricacies of nuclear
interactions. In fact, one of the early projects Gerry assigned to me was
the coupling of $\rho$-mesons to the little bag, which I worked out within
a few weeks and which was supposed to be the starting point for a more
microscopic investigation of vector meson exchanges in the nuclear force, thus
further contributing to the sought-after unified picture of nucleons and 
the nuclear interactions.
When I presented to Gerry my notes in the form of a handwritten draft -
\LaTeX was not yet available - he just told me that I was too fast for him and
proposed to me that I work alone.  So I  teamed up with Andreas Wirzba - a fresh
new graduate student from M\"unster - and Ismail Zahed, then a post-doc in
Gerry's group to work on Casimir effects in chiral bag models and on many
aspects of Skyrmions, which was a hot topic at that time. The
Skyrme model promised to lead to the long wanted unification of nucleon
and nuclear physics, and many impressive calculations revealed further
deep connections, but again the lack of a power counting was the stumbling
stone.

\section{A first step: Studying pion-nucleon scattering in chiral
perturbation theory}

\begin{figure}[t]
\begin{center}
\includegraphics[width=3cm,angle=270]{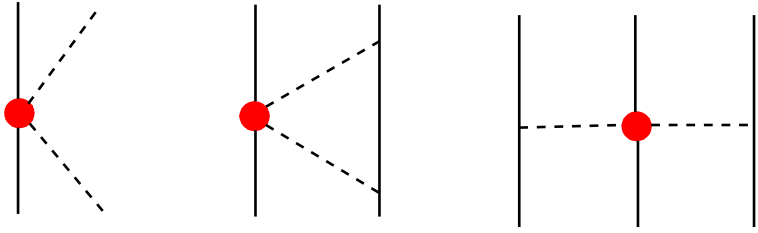}
\end{center}
\caption{\label{fig:ci}
The LECs $c_i$ (circles) in pion-nucleon scattering (left), the 
two-nucleon (NN) interaction (center) and the three-nucleon
(NNN) interaction (right).
}
\end{figure}
\begin{figure}[t]
\begin{center}
\includegraphics[width=7cm,angle=0]{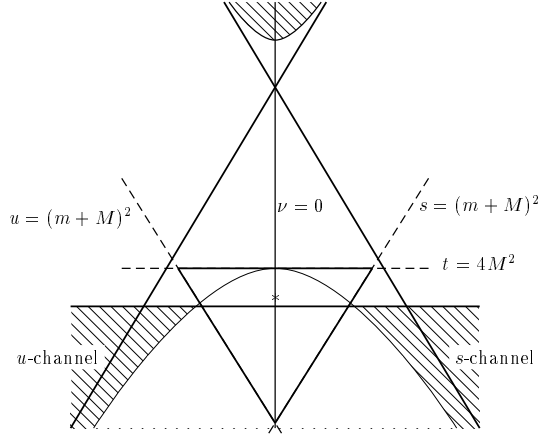}
\end{center}
\caption{\label{fig:mandel}
Mandelstam plane. The Mandelstam triangle is the inside of the
thick lines. 
}
\end{figure}

During my stay as a Heisenberg fellow at Bern, I was lucky to learn
chiral perturbation theory (CHPT) from J\"urg Gasser and Heiri Leutwyler,
who had transformed Weinberg's ideas into an effective machinery to
analyze low-energy QCD. With V\'eronique Bernard and Norbert Kaiser
we systematically analyzed electroweak reactions on the nucleon and
in particular  neutral pion photoproduction off the proton, $\gamma p\to\pi^0 p$, 
triggered by measurements at Saclay, Mainz and Saskatoon to test the then
believed low-energy theorems based on soft-pion algebra (plus one
extra assumption). The explanation of the data by a new one-loop
effect  helped to establish the method as a precision tool
in the single nucleon sector.
In the the late nineties and early years of the new millennium, with 
my students Sven Steininger and Nadia Fettes \cite{piNJ} we investigated 
pion-nucleon scattering in the framework of heavy baryon chiral
perturbation theory - this later turned out to be an important
ingredient in the construction of the few-nucleon forces based
on effective field theory (EFT). The underlying chiral Lagrangian
of pions and nucleons coupled to external currents can be written
as
\beq
{\cal L}_{\rm eff} = {\cal L}^{(1)} + {\cal L}^{(2)}
+ {\cal L}^{(3)} + {\cal L}^{(4)} + \ldots,
\eeq
where the superscript denotes the chiral dimension (the power
in small external momenta and/or pion masses). Tree level
computations are done with ${\cal L}^{(1)} + {\cal L}^{(2)}$,
a complete one-loop calculation involves all loop graphs
with insertions from ${\cal L}^{(1)}$ and  at most one
insertion from  ${\cal L}^{(2)}$ and so on. In its full
glory, ${\cal L}_{\rm eff}$ can be found in~\cite{Fettes:2000gb}.
Of particular importance in these studies are the 
finite dimension two low-energy constants (LECs) $c_i$,
see Fig.~\ref{fig:ci},
that also feature prominently in the description of the two-
and three-nucleon forces - Gerry was one of the first who had
understood this long before the advent of EFT.
An even better
determination of some of the $c_i$ was possible by analytically
continuing the amplitudes into the interior of the Mandelstam
triangle, see Fig.~\ref{fig:mandel} - the ability to use dispersion
relations as I learned it in Stony Brook turned out to be
very valuable. All this work can be summarized by the following
values for the $c_i$ in units of GeV$^{-1}$:
\beq\label{eq:ci}
c_1 = -0.9^{+0.2}_{-0.5}~,~~ c_2 = 3.3 \pm 0.2~,~~ c_3 = -4.7^{+1.2}_{-1.0}~,~~
c_4 = 3.5^{+0.5}_{-0.2}~.
\eeq
\noindent 
The values for $c_2, c_3$ and $c_4$ are larger than the expected (natural)
values: {$c_i \sim g_A/\Lambda_\chi \simeq 1.1 \ldots 1.5$, with $g_A = 1.267$
the nucleon axial-vector coupling and $\Lambda_\chi \simeq 1\,$GeV the
scale of chiral symmetry breaking. This was understood early in terms
of resonance saturation - the fact that the heavier states integrated out
from the EFT leave their imprint in the values of the LECs. This method
was pioneered in the meson sector by Gerhard Ecker and collaborators and
also by John Donoghue and collaborators \cite{reso}. In 1997, with V\'eronique
Bernard and Norbert Kaiser, we had already determined the $c_i$ by a
comparison to scattering lengths and subthreshold parameters and extended
the resonance saturation scheme to the one-baryon sector, see
Fig.~\ref{fig:resosat}. Not surprisingly, the LECs $c_2$ and $c_3$ are
largely dominated by the close-by $\Delta(1232)$ resonance, whereas
$c_4$ also receives an important contribution from the $\rho$-meson \cite{Bernard:1996gq}.
%\noindent 
This leads to the following values of $c_{2,3,4}$ (in brackets the
ranges obtained from resonance saturation) $c_2 = 3.9~[2 \ldots 4],
c_3 = -5.3~[-4.5 \ldots -5.3], c_4 = 3.7~[3.1 \ldots 3.7]$, whereas $c_1$
is given by scalar meson exchange ($\pi\pi$ correlations). These numbers
are also consistent with the recent analysis of pion-nucleon scattering
and subthreshold parameters derived from pionic hydrogen and deuterium,
$c_1 = -1.2 \ldots -0.9, c_2 = 2.6 \ldots 4.0, c_3 = -6.1 
\ldots -4.4$~\cite{Gasser:2007zt}. Let me also stress that the uncertainties
in the $c_i$ are strongly correlated, as best seen from the combination
$-2c_1 + c_2 +c_3 - g_A^2/(8m)$ that determines the small isoscalar pion-nucleon
scattering length.
\vspace{2mm}
\begin{figure}[ht]
\begin{center}
\includegraphics[width=2.5cm,angle=270]{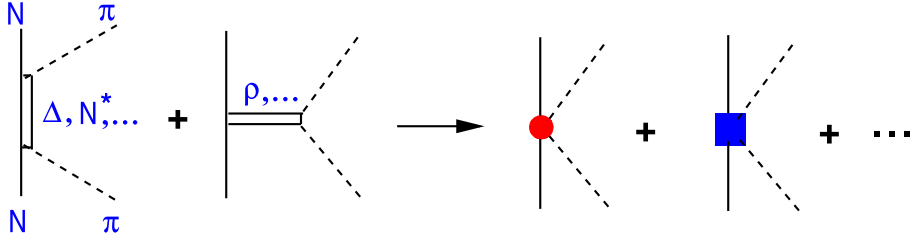}
\end{center}
\caption{\label{fig:resosat}
Resonance saturation for the dimension two (circles) and three (squares)
LECs.
}
\end{figure}

\section{Flashback: The Texas revolution}

Already in the early nineties a decisive step towards a systematic
theory of the nuclear forces had been done. This was initiated by a 
series of papers by Steven Weinberg, who had already pioneered the 
use of effective chiral Lagrangians decades earlier under the nowadays
somewhat strange-reading title ``Phenomenological Lagrangians''
\cite{Weinberg:1978kz}. In that marvelous paper the idea of the
power counting was developed in detail and further a very illuminating
application of the renormalization group to work out the so-called chiral logarithms
was presented. Years later, while teaching the effective field theory of pions and
nucleons, Weinberg suddenly realized that there are additional terms consistent
with the power counting, terms quadrilinear in the nucleon fields\footnote{I
  am grateful to Steve Weinberg for providing me with this recollection.}. He
immediately realized that these must be related to the much discussed 
mechanism for the short-range repulsion of the nuclear forces. Here the
fundamental principle of effective field theory works at it best - in the
low-energy sector of a given theory one is not able to resolve the physics
at large momentum scales but simply parameterizes it in terms of operators made
of the fields one has at one's disposal - it just does not matter whether
the nucleon-nucleon repulsion is due to vector-meson exchange, quark
rearrangement energies or whatever model has been invented before.
Weinberg then 
worked out the consequences of the spontaneous chiral symmetry breaking 
for the forces between two, three and four nucleons using the method
of phenomenological Lagrangians \cite{Weinberg:1990rz}, nowadays called 
chiral perturbation theory or chiral effective field theory. Remarkably, 
the first of these
two papers contains only three references, one to the review Gerry
had written in 1985 with Sven-Olaf B\"ackman and Jouni 
Niskanen~\cite{Backman:1984sx} and also, he explicitly thanks Gerry
for  ``enlightening conversations on nuclear forces''. The 
important step taken by Weinberg was to realize that in the nuclear
force problem one can not apply the power counting directly to the
S-matrix, but rather to the effective potential - these are all diagrams
without $N$-nucleon intermediate states. Such diagrams lead to pinch 
singularities in the infinite nucleon mass limit (the so-called static limit), 
so that e.g. the  nucleon box graph is enhanced as $m/Q^2$, with $m$ the nucleon mass
and $Q \ll \Lambda_\chi$ a small momentum. The beautiful power counting formula for the
graphs contributing with the $\nu$th power of $Q$ or a pion mass  
to the effective potential reads (considering only connected pieces):
\beq
\nu = 2 - N - 2L + \sum_i V_i \left[ d_i + \frac{n_i}{2} -2 \right]~.
\eeq
Here, $N$ is the number of in-coming and out-going nucleons, $L$ the
number of pion loops, $V_i$ counts the vertices of type $i$ with $d_i$
derivatives and/or pion mass insertions and $n_i$ is the number of nucleons
participating in this kind of vertex. Because of chiral symmetry, the
term in the square brackets is larger than or equal to  zero and thus the 
leading terms contributing e.g. to the NN potential can easily be 
identified. These are the time-honored one-pion exchange and
two four-nucleon contact interactions without derivatives. 
These contact interactions were indeed the missing part which
were earlier modeled by heavy meson-exchanges or by some
kind of fit function - but none of these approaches was controlled or
systematic. Here again the awesome power of power counting 
becomes crystal-clear - it just took one person not biased
by nuclear folklore to take this final step.
The so-constructed effective potential is then iterated
in the Schr\"odinger or Lippman-Schwinger equation, generating
the shallow nuclear bound states as well as scattering states.
The resulting contributions to the 2N, the 3N and the 4N forces
are depicted in Fig.~\ref{fig:power}. I will come back to a
detailed discussion of the various entries below, at this
point it is important to observe that - consistent with 
phenomenological observations - three-nucleon forces appear
only two orders after the dominant NN forces and four-nucleon
forces are even further suppressed, appearing only at N$^3$LO.
Let me stress that the beautiful analysis of  few-nucleon forces 
in chiral EFT by Bira van Kolck \cite{vanKolck:1994yi}
was an important ingredient to set up the power counting as
displayed in  Fig.~\ref{fig:power} and
that this extended power counting
has been the major development on the way to the chiral theory of
nuclear forces. 

\begin{figure}[t]
\begin{center}
\includegraphics[width=14.0cm,angle=0]{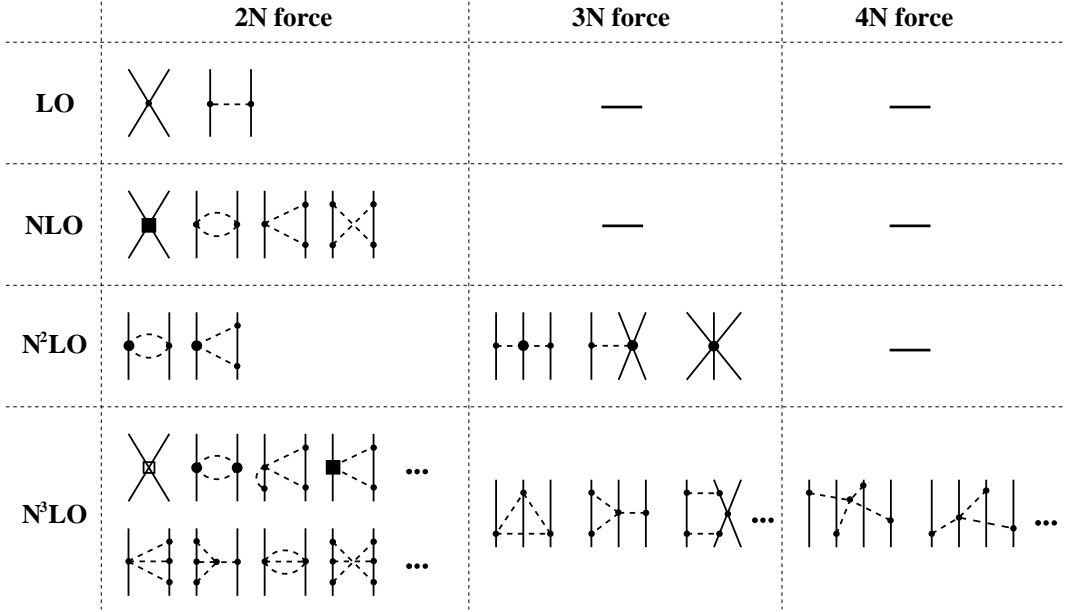}
\end{center}
\caption{\label{fig:power}
Contributions to the effective potential of the 2N, 3N and 4N forces
based on Weinberg's power counting.
Here, LO denotes leading order, NLO next-to-leading order and so on.
Dimension one, two and three pion-nucleon interactions are denoted
by small circles, big circles and filled boxes, respectively. In the
4N contact terms, the filled and open box denote two- and four-derivative
operators, respectively.
}
\end{figure}

The first large-scale numerical evaluation of this chiral potential
was done by Bira van Kolck and collaborators a few years after Weinberg's
seminal papers, see Refs.~\cite{Ordonez:1993tn}. They worked
up-to-and-including N$^2$LO in a theory of pions, nucleons and also
the $\Delta$ isobar, fitting the  LECs of the contact interactions
to low-energy NN phases separately for isospin $I=0$ and $I=1$ and
properties of the deuteron. This work was certainly
ground-breaking but it was also not immediately accepted in the nuclear physics
community - for the simple reason that the phenomenological approaches
to the two-nucleon problem had achieved a much higher precision - at the
expense of many more parameters and lack of  systematicity. Still, the first
important step was done. Before picking up this theme, let me stress that
Weinberg extended his approach to pion reactions on light 
nuclei~\cite{Weinberg:1992yk}, using at that time deuteron wave functions
from phenomenological potentials. This so-called ``hybrid appraoch'' laid 
the ground for systematic studies of pion scattering off light nuclei, 
pion photo- and electroproduction off light nuclei and of pion production
in NN collisions.  

\section{The chiral theory of nuclear forces as a precision tool}

\subsection{The nucleon-nucleon interaction}
When I was still at Bonn filling in for the vacant chair of Max Huber, then 
rector of Bonn University, I met a few times with Karl Holinde of the IKP
J\"ulich. Karl had become famous for his work on the Bonn potential and 
was hired by Josef Speth at J\"ulich in 1984, which gave Karl a fantastic
home base to continue his work. We talked about the chiral potential
of the Texas group, and he was clearly interested but also aware that the
precision was not yet competitive. So we decided to work together to try
to improve the Texas potential. Unfortunately, before really getting started,
this collaboration was terminated by Karl's untimely death in December of 1996.
In fact, Gerry came to J\"ulich the next year and gave a wonderful talk about the
roots, the physics and the achievements of the Bonn potential, focusing on
Karl's contribution. Fortunately, a bit later Walter Gl\"ockle from Bochum
contacted me - he had acquired a brilliant young student named Evgeny
Epelbaum,  who had done some calculations of the chiral NN potential in his 
diploma thesis and was supposed to continue working on this topic for his
doctoral thesis. Walter was (and still is) a world-leading expert in 
few-nucleon calculations but did not feel at
ease with effective chiral Lagrangians. So we decided that we would team
up to supervise Epelbaum, which turned out to be a true pleasure as we
did not have to do much. Being aware of the short-comings of energy-dependent
potentials in few-body calculations, we decided to recalculate the Texas
potential using the method of unitary transformations - which I had not used
for about  15 years. Based on Epelbaum's diploma thesis, we published
a paper on the power counting adapted to this method in~\cite{Epelbaum:1998ka}.
Before working out the phenomenological consequences of this, we decided
to work out some simpler examples to get better acquainted with the many
subtleties of the nuclear interaction and how they reflect on the
corresponding effective theory. Interestingly, in two papers we developed
a method to integrate out the high-momentum components from an effective
theory based on a simple local potential of the Malfliet-Tjon type \cite{Epelbaum:1998hg}. 
This preceeded the work of Achim Schwenk, Tom
Kuo, Bengt Friman, Gerry and others at Stony Brook on the so-called $V_{\rm
  low-k}$ potentials based on elegant renormalization group techniques, see
e.g. the review \cite{Bogner:2003wn} and references therein. Such parallel
developments are a fine witness of Gerry's legacy - he has taught many
generations of students a deep understanding of physics that will necessarily
lead to progress in the field of nuclear physics. But back
to the nuclear force problem. We published our N$^2$LO 
results in~\cite{Epelbaum:1999dj}, which were markedly improved as compared to
the Texas potential. The simple reason was that we did not make global fits 
but rather projected the chiral potential onto partial waves and determined
the four-nucleons LECs by fitting to the S- and P-waves and the
$^3S_1$-$^3D_1$ mixing parameter in the neutron-proton system. An important paper that
had been published earlier was the work  by the Munich group, where
the role of the two-pion exchange to the peripheral partial waves was scrutinized
\cite{Kaiser:1997mw}. They demonstrated that these peripheral waves are well
described by one- and two-pion exchanges in the Born approximation, thus only
the low partial waves were to be used in fitting the multi-nucleon LECs.
However, the large values of the $c_i$,
cf. Eq.(\ref{eq:ci}), lead to a very strong isoscalar central potential at
short distances that generates unphysical deep bound states for the larger
values of the cut-off scale ($\Lambda \simeq 1\,$GeV) in the LS-equation.
Again, earlier work done by Chemtob, Durso and Riska -- that was taught at 
Stony Brook -- came to our rescue,
we simply adopted the so-called spectral function regularization method
to deal with the unwanted short-distance behavior of the two-pion 
exchange \cite{Epelbaum:2003gr}. 
\begin{figure}[tp]
\begin{center}
\includegraphics[width=5.5cm]{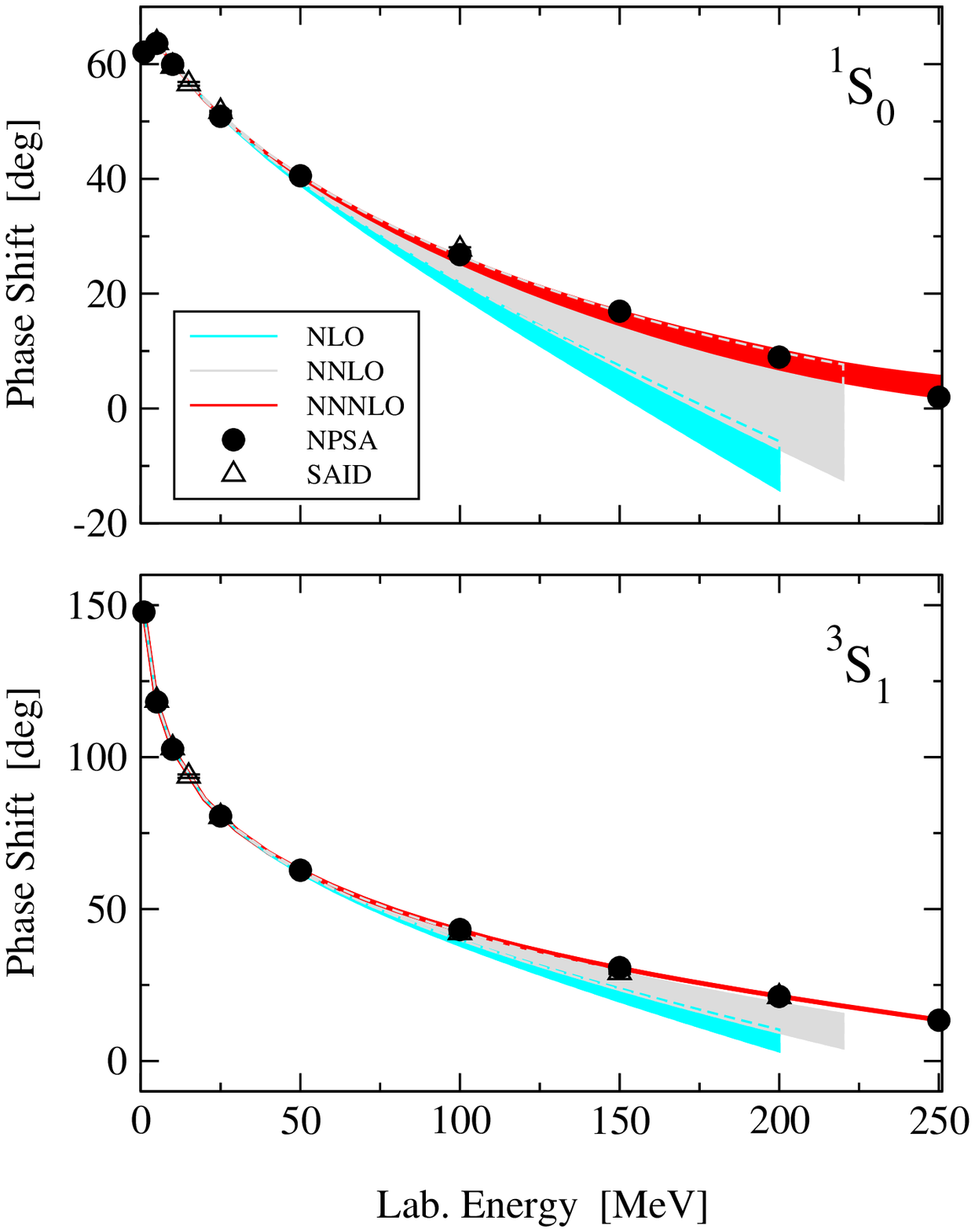}\hspace{0.5cm}
\includegraphics[width=6.25cm]{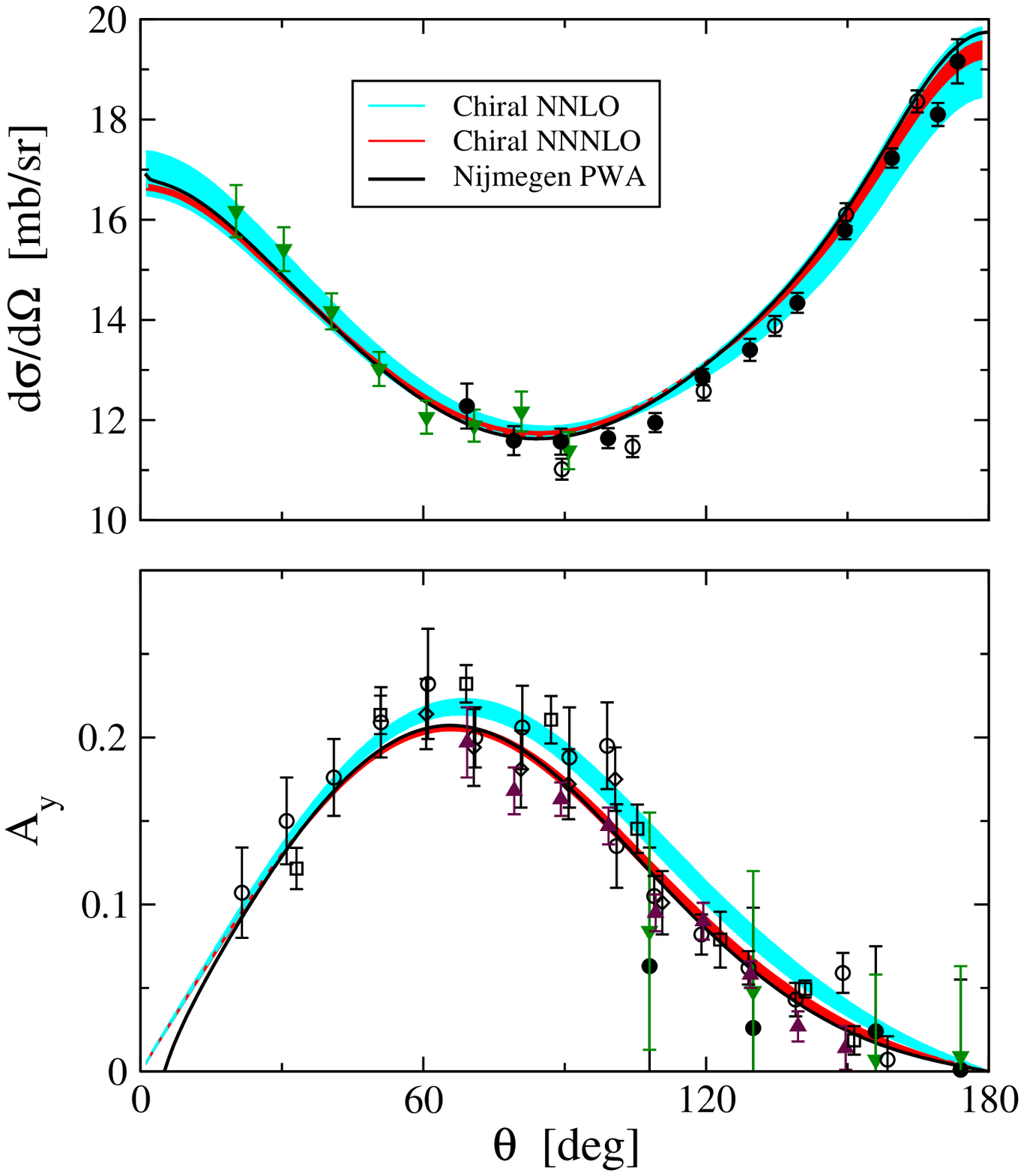}
\end{center}
\caption{\label{fig:2Nn3lo}
Left: $np$ S-waves at  NLO (grid), N$^2$LO (light shaded) and 
N$^3$LO (dark shaded). Right: Differential cross section and vector
analyzing power for $np$ scattering at $E_{\rm lab} =50\,$MeV compared 
to the N$^2$LO (light shaded) and N$^3$LO (dark shaded) predictions.  
}
\end{figure}
\noindent
So the ground was prepared for going to
yet higher orders. While Evgeny had moved to Jefferson Lab as the first
Nathan-Isgur-Distinguished-Fellow, we continued our collaboration and
published the  N$^3$LO results in~\cite{Epelbaum:2004fk}. Earlier on,
Ruprecht Machleidt and David Entem had published  N$^3$LO results using dimensional
regularization in the two-pion exchange \cite{Entem:2001cg}
based on a series of papers from
Norbert Kaiser - this story is told in the contribution of Machleidt to this volume.
Some characteristic results of our N$^3$LO calculation are displayed
in Fig.~\ref{fig:2Nn3lo} - these are  not more accurate than the ones based on
phenomenological potentials, however, they provide a measure of the
theoretical uncertainty and, more importantly, they are based on the chiral 
symmetry of QCD. To close the circle, one can indeed show that the
four-nucleon operators to a large extent are saturated by resonance
excitations, as explicit calculations mapping boson-exchange models onto
the chiral expansion of the effective potential showed, for details
see~\cite{Epelbaum:2001fm}. In a way, EFT provides a reason why the 
meson-exchange models of the nuclear force have been so successful.

\subsection{Three-nucleon interactions}
\label{sec:three}

One of the biggest advantages of the chiral Lagrangian approach to the nuclear
force problem - as stressed early in Weinberg's papers - is the consistent
derivation of the 3- and 4-nucleon forces and also the meson-exchange currents
(that is, the response to electroweak probes). It is worth mentioning that
decades earlier Gerry was aware of the role of chiral symmetry in the
description of three-body forces in nuclei - see his nice paper with
Saul Barshay in 1972 \cite{Barshay:1974aj}. 
But let me come back to the modern approach. As already noted,
the three-nucleon force (3NF) only appears two orders after the leading NN 
interaction. At this order, there are only three topologies contributing,
see Fig.~\ref{fig:3nfdia}.
\begin{figure}[t]
\begin{center}
\includegraphics[width=5.0cm,angle=0]{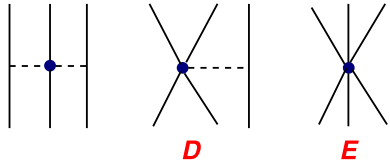}
\end{center}
\vspace{-3mm}
\caption{\label{fig:3nfdia}
Topologies of the leading contributions to the chiral 3NF. From left to
right: Two-pion exchange, one-pion-exchange and 6N contact interaction.
}
\end{figure}
\noindent The two-pion exchange topology is given again in terms of the $c_i$,
as  discussed in detail in~\cite{Friar:1998zt}. The so-called $D$-term, which 
is related to the one-pion exchange between a 4N contact term and a further
nucleon, has gained some prominence over the last years, as many authors have
tried to pin it down based on a cornucopia of reactions, such as $Nd \to Nd$,
$NN \to NN\pi$, $NN\to d \ell \nu_\ell$ or $d \pi \to \gamma NN$.
This demonstrates nicely the power of EFT -
very different processes are related through the same LECs thus providing many
different tests of chiral symmetry (as it is also the case with the
LECs $c_i$, see Fig.~\ref{fig:ci}). The LEC $E$ related to the 6N contact 
interaction can only be fixed in systems with at least three nucleons, say
from the triton binding energy. Remarkably, although there are many diagrams at
N$^3$LO, there are no new LECs to be determined. The long-range terms of this
force can be found in~\cite{Bernard:2007sp} and the shorter ranged ones will
be published soon (see also \cite{Ishikawa:2007zz}). 
Many applications of these forces and the testing of their
structure can be found in the reviews \cite{Epelbaum:2005pn,Epelbaum:2008ga}.
%
%\begin{figure}[t]
%\begin{center}
%\includegraphics[width=8.0cm,angle=0]{Dterm.eps}
%\end{center}
%\vspace{-3mm}
%\caption{\label{fig:D}
%Appearance of the $D$-term in various low-energy reactions.
%}
%\end{figure}
%

As a very nice example that these forces make their way into nuclear structure
calculations, I show here the results from Petr Navr\'atil and collaborators 
\cite{Navratil:2007we}, who performed large-basis no-core shell model (NCSM)
calculations including the leading 
chiral 3N forces and demonstrated their necessity to describe the spectra
of nuclei with $A = 10, \ldots 13$, see the entry NN+NNN
in  Fig.~\ref{fig:ncsm}. Note, however, that in this calculation the 2N force
was employed at N$^3$LO while the 3NF was only included at N$^2$LO. It will
be interesting to see how the inclusion of the sub-leading 3NF terms will
modify these results.
\begin{figure}[t]
\begin{center}
\hspace{-7.3cm}
\includegraphics[width=7.5cm,angle=0]{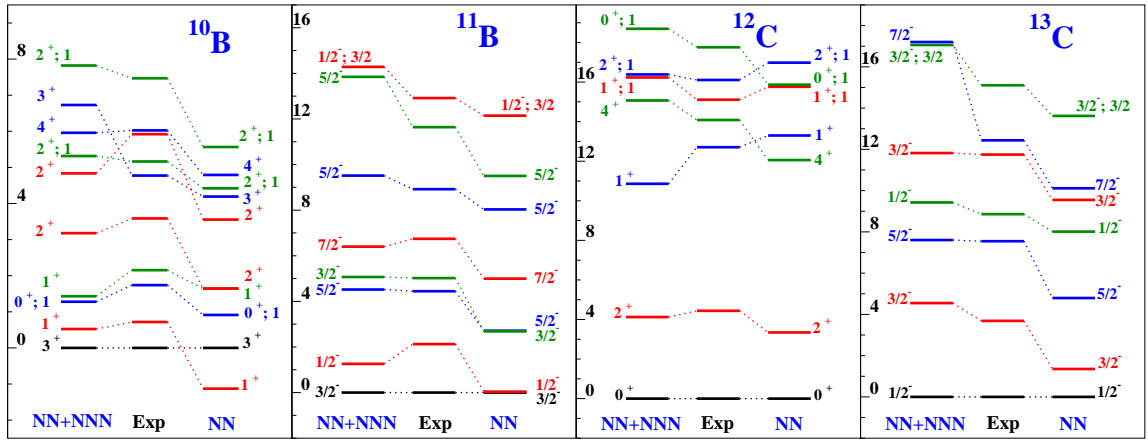}
\end{center}
\vspace{-3mm}
\caption{\label{fig:ncsm}
Spectra of light nuclei using chiral NN and chiral NN+NNN forces in a NCSM 
calculation compared to the data. Figure adopted from \cite{Navratil:2007we}
and courtesy of P.~Navr\'atil.
}
\end{figure}

So far, I have only discussed the EFT with pions and nucleons. Clearly, the
two-pion exchange diagram in Fig.~\ref{fig:3nfdia} is a controlled
approximation to the time-honored  Fujita-Miyazawa force, whose 50$^{\rm th}$
anniversary was celebrated in Tokyo in 2007. For a light reading of how the
Fujita-Miyazawa force is related to the EFT description of the 3NF, I refer
to \cite{Meissner:2008zza} and references therein.

\subsection{The nuclear matter problem}
\label{sec:nucmat}

Another important question in nuclear physics is to understand the 
saturation properties of nuclear matter - an idealized infinite
system of nucleons in which all Coulomb effects are switched off.
From the properties of heavy nuclei using some sophisticated mass
formula, one can extrapolate to nuclear matter - and determine its
saturation properties. The binding energy per nucleon $E/A$ is
approximately $-16\,$ MeV at a Fermi momentum of about 1.3~fm$^{-1}$.
So what does chiral EFT have to say? A first important step was taken
by Norbert Kaiser, Wolfram Weise and collaborators at M\"unchen, 
who calculated the contribution of pion exchange(s) to the energy
density of nuclear matter and showed that the energy density of 
isospin symmetric nuclear
matter can be extremely well approximated by the simple form
\beq
E/A = \frac{3k_F^2}{10m} - \alpha \frac{k_F^3}{m^2} + \beta
\frac{k_F^4}{m^3}~.
\eeq 
They then calculated the coefficients $\alpha$ and $\beta$ from chiral 
dynamics. With some fine-tuning of the  regulator, one
finds an astonishingly good description of the energy density of
nuclear matter \cite{Kaiser:2001jx}. This was later improved by
including e.g. higher orders (sensitive again to the LECs $c_i$)
and isobar degrees of freedom \cite{Fritsch:2004nx}. Interestingly,
all this was based on a loop expansion with no explicit power 
counting - a power counting that indeed explained the success of
these calculations was only set up much later, see \cite{Oller:2009zt}.
Space forbids a discussion of this power counting and the resulting
physics in detail - I only would like to mention that it could be
shown that for many reactions the contributions from the multi-nucleon
interactions cancel to leading order which is at the heart of the
success of the Munich group calculations.

Here, I briefly pursue a more ``conventional'' approach to the nuclear
matter problem based on more standard nuclear many-body theory. For decades,
the method of choice to analyze nuclear matter has been the G-matrix
of Brueckner and others - leading to the problem that for phase-shift
equivalent NN potentials the resulting  saturation properties lie
on the Coester line that does not pass the empirical values. Recently,
Siegfried Krewald and collaborators have recalculated the properties of 
nuclear matter based on the chiral 
NN and 3N forces (see the v2 version of Ref.~\cite{Saviankou:2008tg})
utilizing the R-matrix as proposed by Baker in 1971
\cite{Baker:1971vm}. As can be seen from the left panel of
Fig.~\ref{fig:nucmat}, the NN interaction alone  only binds
for Fermi momenta larger than 1.7~fm$^{-1}$. However, including the
3NF with natural values for $D$ and $E$, one finds binding with the
proper strength at the proper density. In the right panel, the
sensitivity to the two cut-offs in the NN force as compared to
phenomenological determinations is shown, it is of comparable size.
The dependence on the 3-body LECs $D$ and $E$ is also weak, see
Fig.~3 in Ref.~\cite{Saviankou:2008tg}. It will be interesting to push these
calculations to N$^3$LO and to include the $\Delta(1232)$ in the EFT.

\begin{figure}[t]
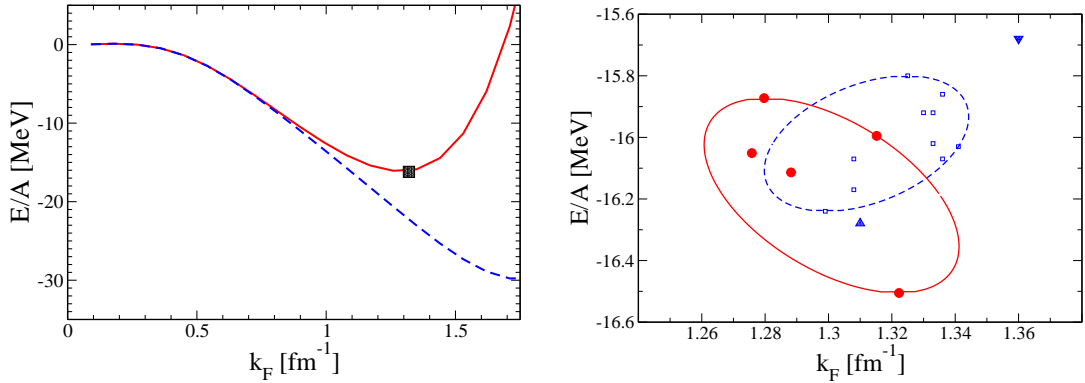

\begin{center}
%\hspace{-7.3cm}
\includegraphics[width=6.75cm,angle=0]{ea103wsq.eps}\hspace{0.5cm}
\includegraphics[width=6.75cm,angle=0]{sat_ellipse2.eps}
\end{center}
\vspace{-3mm}
\caption{\label{fig:nucmat}
Left panel: Binding energy per particle of nuclear matter as a function of the
Fermi momentum $k_F$ for the N$^2$LO potential using
a Lippman-Schwinger cut off $\Lambda = 550$~MeV and a spectral function
cut off $\tilde{\Lambda} = 600$~MeV (solid line).
The binding energies obtained in the absence of three-nucleon interactions
are shown by the dashed line. The black square gives the empirical nuclear
matter properties. Right panel: 
Saturation points of nuclear matter.
Downward triangle, upward triangle and rectangle: various phenomenological
approaches. Circles: chiral EFT at N$^2$LO.
}
\end{figure}

\subsection{Back to square one - meson-exchange currents}

\begin{figure}[tp]
\begin{center}
\includegraphics[width=13.0cm,angle=0]{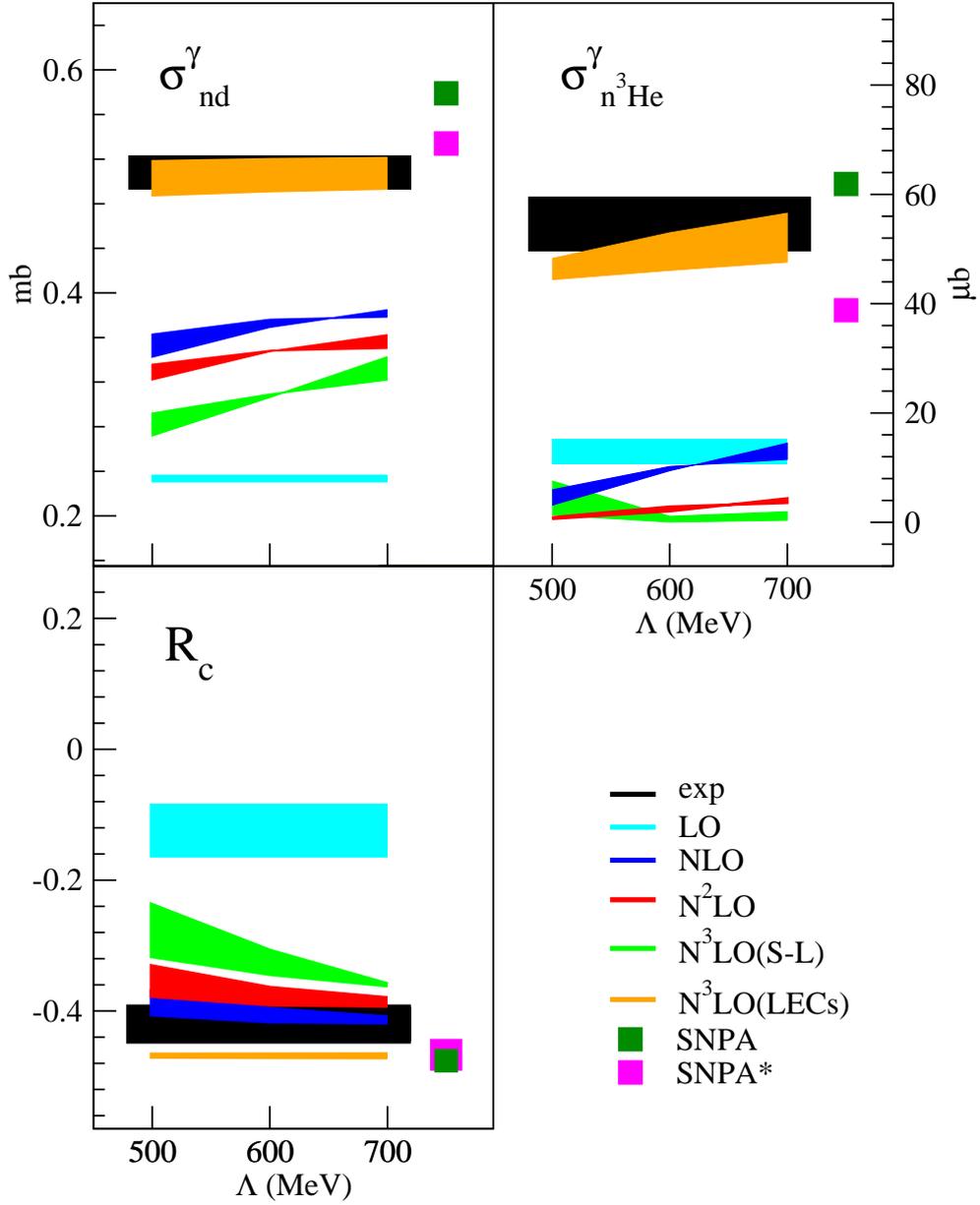}
\end{center}
\vspace{4mm}
\caption{\label{fig:MEX}
Predictions for the $n$-$d$ and $n$-$^3He$ capture reactions.
Chiral EFT: the curve labeled N3LO(S-L)
includes pion loop corrections  contact terms, 
while the curve labeled N3LO(LECs) includes in addition
contributions from (higher order) pion exchanges
and non-minimal contact currents. Pre-EFT approaches: 
SNPA and SNPA* denote up-to-date conventional calculations.
Figure courtesy of Rocco Schiavilla.
}
\end{figure}
As told in section~1, I started my career calculating meson-exchange
currents based on the method of unitary transformations. It is quite
a nice turn of events that in chiral nuclear EFT one is now able to do
this in a controlled and systematic manner, thanks to the power counting.
In our approach, Stefan K\"olling, Hermann Krebs and Evgeny Epelbaum
have taken up the task -  the two-pion exchange electromagnetic current
based on this method is published in  \cite{Kolling:2009iq} and together
with the Cracow group led by Henryk Witala and Jacek Golak a thorough
investigation of electro-nuclear processes is underway, for a first
application to deuteron photodisintegration see~\cite{Rozpedzik:2010kc}.
However, the group around Rocco Schiavilla, using old-fashioned time-ordered
perturbation theory, has taken the lead and performed calculations
of the MECs of one- and two-pion range and applied this to magnetic
moments of light nuclei~\cite{Pastore:2008ui}. 
Most recently, they have calculated thermal neutron capture on $d$ and $^3$He
\cite{Girlanda:2010vm}. The resulting radiative capture cross sections 
$\sigma_{nd}^\gamma$ and $\sigma_{n ^3He}^\gamma$ and the photon circular 
polarization parameter $R_c$, resulting from the capture of polarized neutrons
on the deuteron, are shown in Fig.~\ref{fig:MEX} in comparison to two
state-of-the-art conventional approaches. As the authors note,
these processes are not the best or simplest to illustrate the 
convergence pattern of chiral nuclear EFT. So much more work is needed 
here and will be done.

\section{Flash forward: Combining nuclear EFT with lattice simulations}

\begin{figure}[t]
%\begin{center}
\hfill\includegraphics[width=0.60\textwidth,angle=0]{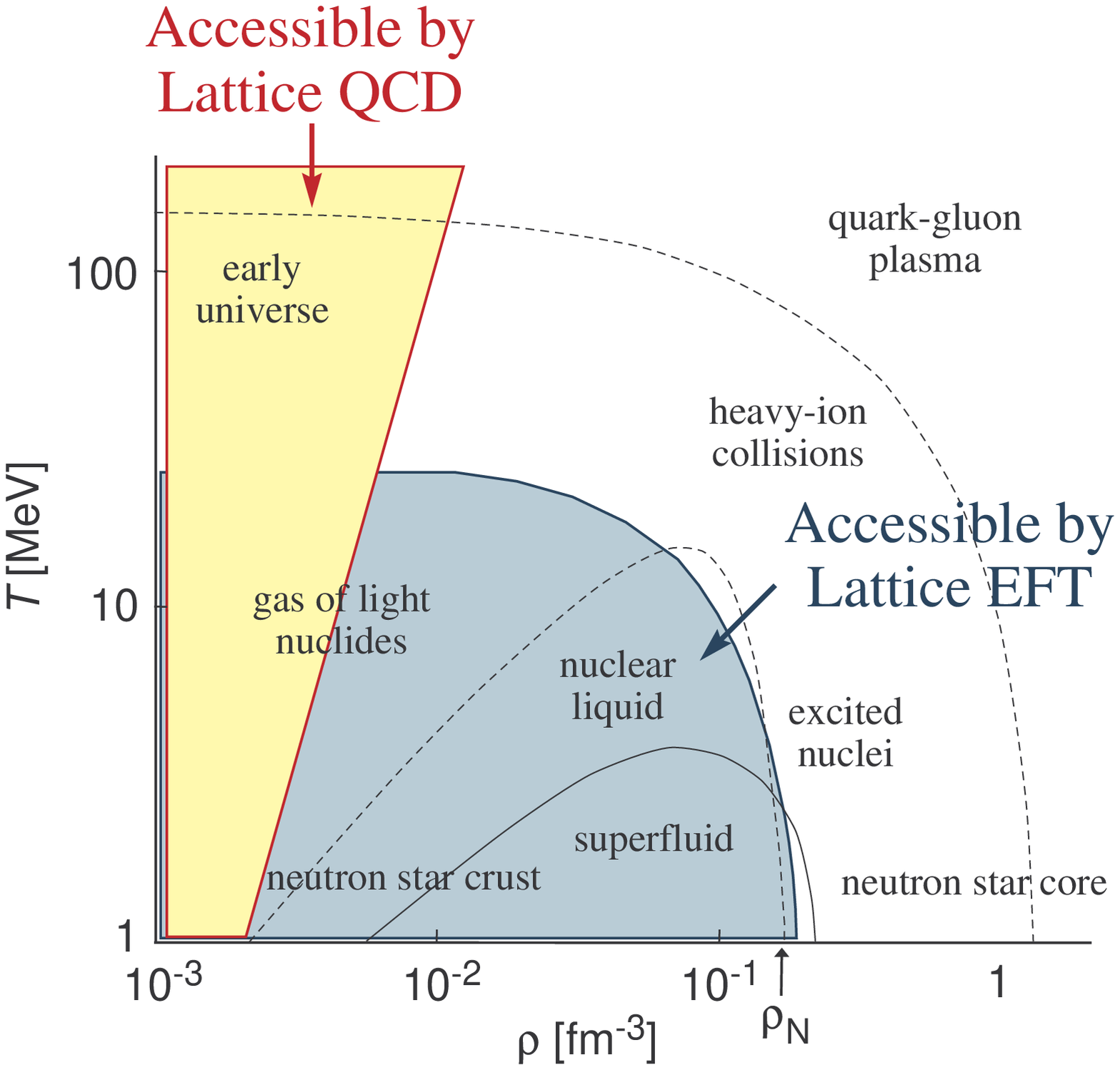}

\vspace{-7.0cm}
\hspace{0.0cm}

\includegraphics[width=0.35\textwidth,angle=0]{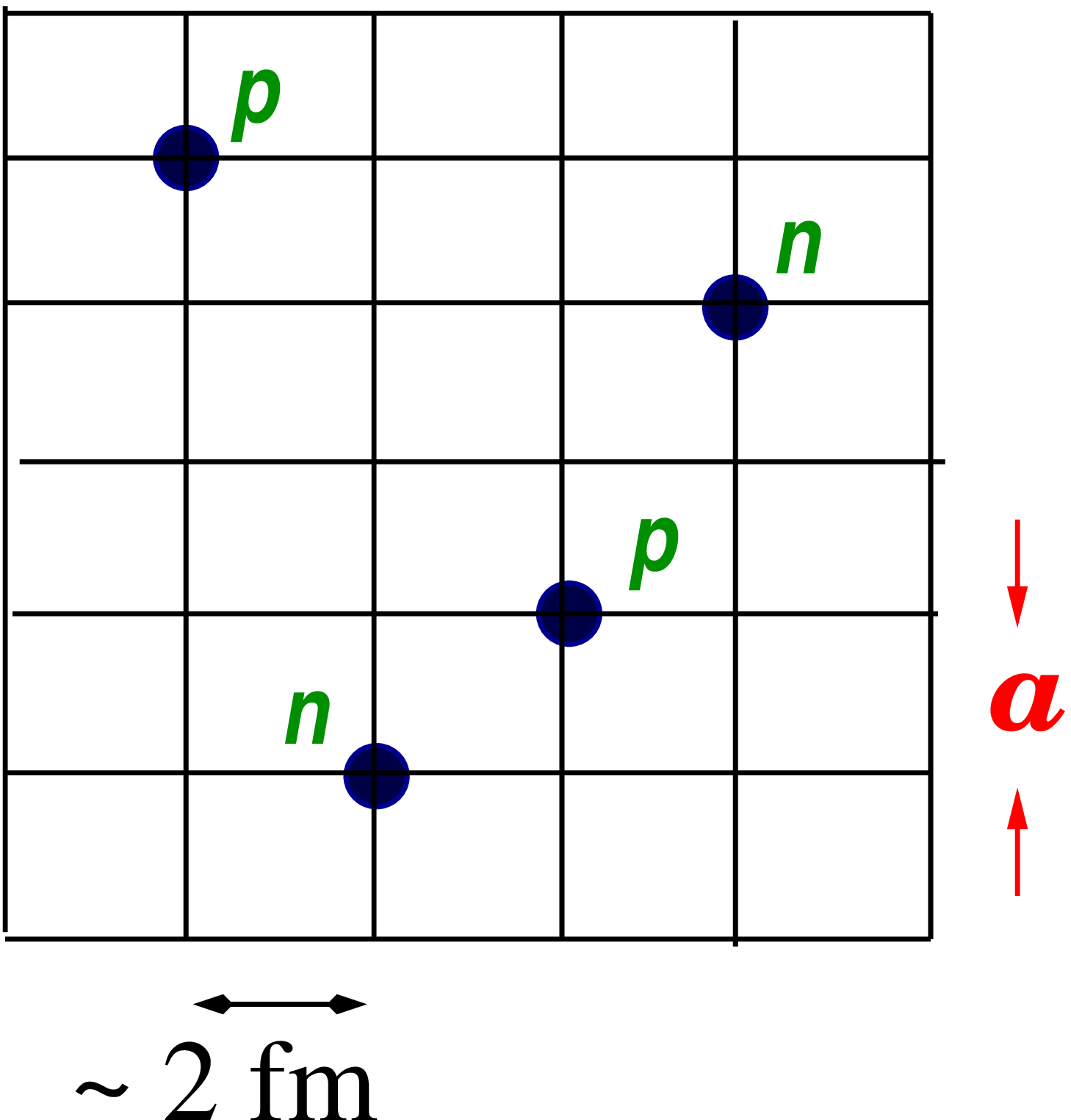}
%\end{center}
\vspace{5mm}
\caption{\label{fig:nuclat}
Left panel: Euclidean space-time lattice with point-like nucleons
on the lattice sites and a typical lattice spacing of $a =  2$~fm.
Right panel: Nuclear phase diagram as accessible by lattice QCD
and by nuclear lattice EFT. Figure courtesy of Dean Lee.
}
\end{figure}

Having now constructed nuclear forces between two, three and four nucleons,
one has a new tool to address the nuclear many-body problem. One venue is
to combine standard many-body techniques with these forces, as exemplified
by the NCSM calculation shown in  Sec.~\ref{sec:three} or the discussion
of nuclear matter in Sec.~\ref{sec:nucmat} (see also the contribution by
Josef Speth, Siegfried Krewald and Frank Gr\"ummer to this Festschrift).
Here, I briefly want to outline a very different and novel approach that
combines the chiral EFT for the forces with the method of Monte Carlo
simulations, that are so successfully utilized in the lattice approach to
QCD. The basic idea is to formulate the chiral EFT on a Euclidean space-time
lattice as depicted in the left panel of Fig.~\ref{fig:nuclat} - here, the
lattice spacing serves as an UV regulator and has to be chosen such that
the nucleons - which are treated as non-relativistic point-like particles
on the lattice sites - do not overlap. A lattice spacing of $a = 2\,$fm
entails an UV cut-off $\Lambda = \pi/a \simeq 300\,$MeV. The pion and nucleon
propagators as well as the one- and two-pion exchanges and the multi-nucleon 
contact interactions are written in terms of lattice variables, making use
of Hubbard-Stratonovich fields to bring the 4- and 6-nucleon interactions
into a quadratic form (for more details, see \cite{Borasoy:2006qn} and 
references therein as well as the nice review by Dean Lee~\cite{Lee:2008fa}). 
Given this Euclidean formulation of the action, the generating functional 
of the theory can then be evaluated with stochastic methods, such as the hybrid 
Monte Carlo approach. It is important to stress that the approximate Wigner 
SU(4) symmetry of the nuclear interactions strongly suppresses the sign
oscillations and thus makes such nuclear lattice
simulations easier to handle than lattice QCD calculations. Within this
scheme, a systematic study of nuclei up to $A \simeq 40$ will be possible
using petascale computing as long as the total spin and isospin of the
nucleus under consideration is zero. Also, the nuclear phase diagram
can be studied for a wide range of temperatures and densities, as indicated
in the right panel of Fig.~\ref{fig:nuclat}.
\begin{figure}[t]
\begin{center}
%\vspace{0.0cm}
\includegraphics[width=11.0cm,keepaspectratio,angle=0,clip]{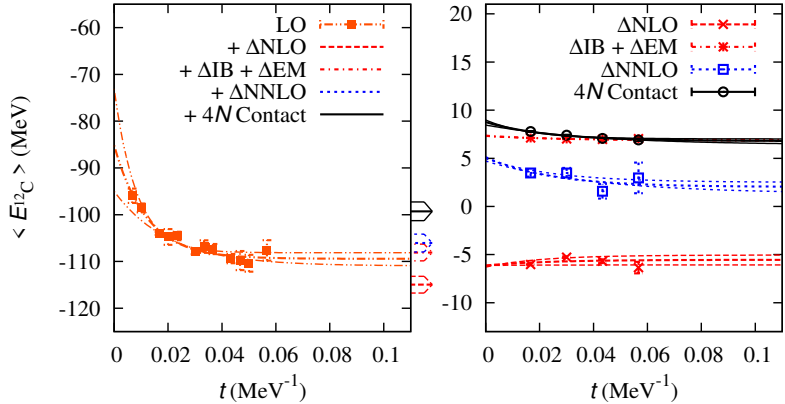}
%\centerline{
%\psfig{file=S_P_D.ps,width=18.0cm}
%}
\vspace{-0.25cm}
\caption[timesteps]{\label{fig:c12} Ground-state energy of  $^{12}$C as a
function of  Euclidean time. For details, see~\cite{Epelbaum:2009pd}.}
\end{center}
\vspace{-5mm}
\end{figure}
The state-of-the-art of these simulations is the calculation of the 
ground-state energy of  $^6$Li and $^{12}$C~\cite{Epelbaum:2009pd}. 
In these works, for the first time the contributions from the Coulomb
interaction between protons and strong isospin-breaking effects were
included. A parameter-free prediction for the energy difference between
the triton and $^3$He could be given,
\begin{equation}
E(^3{\rm H}) - E(^3{\rm He}) = 0.78(5)~{\rm MeV}~,
\end{equation}
in good agreement with the empirical value of  $0.76$~MeV.
The ground-state energies of  $^6$Li and $^{12}$C are calculated
as  $-32.9(9)$~MeV and $-99(2)$~MeV, respectively, not far from
the empirical values of  $-32.0$ and $-92.2$~MeV, cf. also Fig.~\ref{fig:c12}.
This accuracy is comparable to other so-called ab initio calculations
(like the NCSM or Greens function Monte Carlo), that often are based
on a less consistent formulation of the underlying nuclear forces.

Taking  $^{12}$C as the benchmark, it is interesting to estimate the required
computing time and storage needed for the calculation of larger 
nuclei\footnote{I am grateful to Dean Lee for providing me with these estimates.}.
With an improved algorithm, the required CPU time for the  $^{12}$C
simulation on a BlueGene/P architecture and the necessary storage to make
the configurations (about 3500) available is
\beq
X_{^{12}C}^{\rm CPU} = 5 \times 10^{-5}~{\rm PFlop-yr}~,~~~~
X_{^{12}C}^{\rm storage} = 0.14~{\rm TB}~.
\eeq
For a nucleus with $A$ nucleons and with total spin $S$ and isospin $I$
the required CPU time and data storage place can be estimated as
\beqa
X^{\rm CPU} &\approx& X_{^{12}C}^{\rm CPU} \times \left(
  \frac{A}{12}\right)^{3.2}\exp[0.10(A-12) + 3(S~{\rm mod}~2) +4I]~,
\nonumber \\
X^{\rm storage} &\approx& X_{^{12}C}^{\rm storage} \times \left(
  \frac{A}{12}\right)^{2}\exp[0.10(A-12) + 3(S~{\rm mod}~2) +4I]~,
\eeqa
so that e.g. the calculation of the ground-states energies of the magnesium 
isotopes $^{24,25,26}$Mg would require $0.002, 0.064, 0.131$ PFlop-yr and
one would need $1.9, 73.8, 145.5$ TB to store the configurations for the calculation of
matrix elements for these nuclei. In many cases, neutron scattering off nuclei
can also be calculated making use of L\"uscher's formula that relates the
continuum phase shift to the finite energy shift measured on the lattice.
The CPU and storage requirements for neutron scattering of the isotopes $^{24,26}$Mg 
would be  $0.064, 5.43$ PFlop-yr and  $73.8,  5473$~TB, respectively. 
All this work remains to be done but bears a lot of promise. The possibilities 
that would be offered in a future exascale era are
simply breath-taking. 

\section{Some final words}

The chiral effective Lagrangian of QCD offers a tool to systematically
investigate the structure of the nucleon and the nuclear interactions with 
high accuracy - it appears that we are now on the right track to achieve 
what could only be dreamed about a few decades ago - see the discussion in
Sec.~1. Nuclear physics will no longer be based on model-building and phenomenological
approaches, although still quite a bit of work is ahead of us. But this
work - even if hard and time-consuming - will be very rewarding
at the end. Furthermore, strangeness nuclear physics can also be addressed
within this framework \cite{SNP}, but here a sound data basis has yet to be established.
Eventually, lattice QCD might also contribute significantly - but there
are still quite a few obstacles to be overcome, see e.g~\cite{Beane:2009mb}.
It is fair to say that  nuclear physics as a field, and I personally,
owe a lot to Gerry - thank you.

%\vfill
%\pagebreak

\subsection*{Acknowledgements}

I am very much indebted to Gerry - as a physicist and a human.
I am grateful to Evgeny Epelbaum, Dean Lee, Mannque Rho, Rocco Schiavilla,  
and Steven Weinberg for useful communications during the preparation 
of this contribution. I also thank Silas Beane and V\'eronique Bernard for
a careful reading of the manuscript.

%\pagebreak

\end{document}